\documentclass{aastex631}

\usepackage{soul}

\begin{document}

\title{Investigation of Phase Shift and Travel Time of Acoustic Waves in the Lower Solar Atmosphere Using Multi-Height Velocities}

\author[0000-0001-9631-1230]{Hirdesh Kumar}
\affiliation{Udaipur Solar Observatory, Physical Research Laboratory, Dewali, Badi Road, Udaipur 313004 Rajasthan, India.}

\author{Brajesh Kumar}
\affiliation{Udaipur Solar Observatory, Physical Research Laboratory, Dewali, Badi Road, Udaipur 313004 Rajasthan, India.}

\author{Shibu K. Mathew}
\affiliation{Udaipur Solar Observatory, Physical Research Laboratory, Dewali, Badi Road, Udaipur 313004 Rajasthan, India.}

\author{A. Raja Bayanna}
\affiliation{Udaipur Solar Observatory, Physical Research Laboratory, Dewali, Badi Road, Udaipur 313004 Rajasthan, India.}

\author{S. P. Rajaguru}
\affiliation{Indian Institute of Astrophysics, II Block, Koramangala, Bengaluru 560 034, India.}







\begin{abstract}

We report and discuss phase-shift and phase travel time of low-frequency ($\nu$ $<$ 5.0 mHz) acoustic waves estimated within the photosphere and photosphere-chromosphere interface regions, utilizing multi-height velocities in the quiet Sun. The bisector method has been employed to estimate seven height velocities in the photosphere within the Fe I 6173 {\AA} line scan, while nine height velocities are estimated from the chromospheric Ca II 8542 {\AA} line scan observations obtained from the Narrow Band Imager instrument installed with the Multi-Application Solar Telescope operational at the Udaipur Solar Observatory, India. Utilizing fast Fourier transform at each pixel over the full field-of-view, phase shift and coherence have been estimated. The frequency and height-dependent phase shift integrated over the regions having an absolute line-of-sight magnetic field of less than 10 G indicates the non-evanescent nature of low-frequency acoustic waves within the photosphere and photosphere-chromosphere interface regions. Phase travel time estimated within the photosphere shows non-zero values, aligning with previous simulations and observations. Further, we report that the non-evanescent nature persists beyond the photosphere, encompassing the photospheric-chromospheric height range. We discuss possible factors contributing to the non-evanescent nature of low-frequency acoustic waves. Additionally, our observations reveal a downward propagation of high-frequency acoustic waves indicating the refraction from higher layers in the solar atmosphere. This study contributes valuable insights into the understanding of the complex dynamics of acoustic waves within different lower solar atmospheric layers, shedding light on the non-evanescent nature and downward propagation of the acoustic waves.
\end{abstract}



\section{Introduction} \label{sec:intro}

\noindent In 1962, \cite{1962ApJ...135..474L} discovered velocity oscillations on the photosphere of the Sun, depicting a dominant periodicity of the order of 5 minutes. These observed oscillations were theoretically explained by \cite{1970ApJ...162..993U}, and \cite{1971ApL.....7..191L} as standing acoustic waves trapped in the acoustic cavities below the photosphere. \cite{1970ApJ...162..993U} further suggested that these standing waves may exist only along discrete lines in the diagnostic diagram of horizontal wavenumber ($k_{h}$) versus frequency ($\nu$). \cite{1975A&A....44..371D} observationally confirmed the theoretical prediction of the ridges in $k_{h}$ vs $\nu$ diagnostic diagram, and that finally formed the basis for the development of helioseismology, which revealed the dynamics and internal structures of the Sun utilizing the observed oscillations on the photosphere \citep{2002RvMP...74.1073C}. We can infer the vast properties and dynamics inside the Sun through global helioseismology. In contrast, local helioseismology allows us to detect inhomogeneities in the sub-surface layers by measuring the travel times of acoustic waves propagating inside the Sun \citep{2008ASTRA...4...13D}.

Meridional circulation refers to the north-south flow of solar material on the Sun, which is a key component in providing an explanation of the solar cycle \citep{1961ApJ...133..572B, 2015LRSP...12....4H}. Various methods have been utilized to measure meridional flows \citep{1993SoPh..147..207K, 1996ApJ...460.1027H, 2018A&A...611A..92R}; nevertheless, there is significant uncertainty in estimating the deep circulation leading to unclear pictures \citep{2013ApJ...778L..38S, 2013ApJ...774L..29Z, 2015ApJ...805..133J, 2015ApJ...813..114R, 2017PhDT.......153B, 2017ApJ...849..144C, 2020Sci...368.1469G}. Understanding of the deep meridional circulations in the Sun is based on helioseismic measurements. With the help of helioseismic inversion techniques, information about the velocity distribution at different depths below the solar surface can be derived (\cite{2002RvMP...74.1073C, 2005LRSP....2....6G} and references therein). In addition to the challenges posed by the intricate nature of the inversion techniques, particularly when applied to active regions (\cite{2012SoPh..279..323K} and references therein), the reliability of helioseismic measurements itself is impeded by inherent uncertainties and systematic effects. \cite{2003ESASP.517..259D} highlighted a substantial decrease in travel times with increasing latitude for waves that had covered the same travel distances within the Sun. The notable finding of \cite{2009ASPC..416..103D} was the recognition that a systematic effect was impacting the measurements, specifically along the equator, during the examination of asymmetries between wave travel times in opposite travelling directions. Utilizing the time-distance helioseismic analysis technique, \cite{2012ApJ...749L...5Z} found systematic centre-to-limb variation in helioseismic travel times, and suggested that it should be taken into consideration while determining the solar internal meridional flows, whereas the underlying cause of this systematic effect was not well understood. The analysis performed by \cite{2016SoPh..291..731Z} to examine the role of foreshortening in the context of their role in affecting the measurements of acoustic travel times through the time distance approach reveals that foreshortening is not responsible for the systematic centre-to-limb effect in the measured acoustic travel-time differences. Further, they suggested that the formation height of the spectral line, different viewing angles for the convective cells at different disk locations, and maybe other factors are probably among the main causes of the centre-to-limb effect in many helioseismic analyses \citep{2016SoPh..291..731Z}. 
The study by \cite{2018ApJ...853..161C} revealed a remarkable variation in the centre-to-limb effect with frequency. The effect changes its sign at a frequency near 5.4 mHz, close to the disk centre, and reaches its maximum at around 4.0 mHz before the sign reversal. \cite{2020ASSP...57..123Z} indicated that asymmetries found in the computation of travel time in the case of centre-to-limb variation and in the magnetized regions are possibly due to different atmospheric heights, where oscillatory signals are observed for estimating the acoustic travel times. Recently, \cite{2022ApJ...933..109Z} utilizing the long duration multi-height velocities estimated within Fe I 6173 {\AA} line scan observations obtained from the Interferometric BI-dimensional Spectrometer (IBIS; \cite{2006SoPh..236..415C}) installed at the Dunn Solar Telescope (DST) at Sacramento Peak, New Mexico, estimated the phase shift and travel time in the quiet-Sun region, surrounding a small sunspot, of the evanescent acoustic waves. It is to be noted that the helioseismic waves observed in the photosphere are evanescent waves in the quiet Sun. The acoustic cut-off frequency of the quiet-Sun photosphere is around 5.0 mHz \citep{2011ApJ...743...99J}, implying that waves of frequency less than the cut-off frequency do not propagate into the higher atmosphere even the strongest oscillations near the 3.0 mHz. While the acoustic waves above 5.0 mHz propagate into the higher atmosphere with increasing amplitude. It is generally expected that the phase of evanescent waves does not change with height in the atmosphere. The phase shift and travel time analysis performed by \cite{2022ApJ...933..109Z} reveal that the acoustic waves ($<$ 5.0 mHz) carry unexpected phase shift resulting in travel time of the order of around 1 s in the quiet-Sun, suggesting that it may help to understand the centre-to-limb variation of the travel time, which should be taken into account while deriving the inferences of the deep meridional flows. They have also observed a change in a sign around 3.0 mHz in the phase, having the presence of sparse magnetic fields and moat flow around the sunspot; the influence of the same was not assessed in the analysis. \cite{2023ApJ...949...99W} investigated phase shift and phase travel time of evanescent waves in the quiet-Sun, using 3D radiative simulations and synthesized Fe I 6173 {\AA} spectral line profiles. In the simulation, they also found a non-zero phase shift of acoustic waves and reported a substantial difference between the phase measured from Doppler velocity and true velocity. Even though they have also found a non-zero phase shift between simulated velocities for acoustic waves $\nu < $5 mHz, which is quite unexpected. The trend of phase shift vs frequency obtained from true and Doppler velocity is qualitatively similar to the phase shift derived from the IBIS observations. However, the important point is that in the simulation data, sign reversal occurs around 4 mHz, whereas it occurs near 3 mHz in the observational data. Therefore, it is important to understand the phase shift and phase travel time of evanescent acoustic waves within the photosphere of the quiet-Sun, having negligible magnetic fields utilizing multi-height velocities for a better understanding of the physical causes of non-evanescent nature of the low-frequency acoustic waves. Examining the phase shift of the evanescent waves from the photosphere to the chromosphere is imperative, emphasizing the variation of phase shift into the higher solar atmosphere.\\

In this article, utilizing multi-height velocities estimated from the photospheric Fe I 6173 {\AA} and chromospheric Ca II 8542 {\AA} lines over a quiet Sun region in the disk centre, we present an analysis of the phase shift measurement of the evanescent acoustic waves, to better understand the non-evanescent nature of these waves within the photosphere, and photosphere-chromosphere interface regions and to further rule out any possible role of sparse magnetic fields and flows around the sunspot, which was a matter of concern in the earlier analysis. This article is organized as follows: first, we present observational data and reduction in the Section 2, followed by analysis and results in the Section 3. The discussion and conclusions of our results are highlighted in the Section 4.

\section{The Observational Data}
\noindent We have utilized data obtained from the Multi-Application Solar Telescope (MAST: \cite{2009ASPC..405..461M, 2017CSci..113..686V}) operational at the Island observatory in the middle of Lake Fatehasagar of Udaipur Solar Observatory (USO), Udaipur, Rajasthan, India, and space-based Solar Dynamics Observatory (SDO: \cite{2012SoPh..275....3P}) spacecraft to carry out the analysis. First, we provide observational details of the ground-based telescope, followed by the observations taken from the spacecraft.\\  

\subsection{MAST Data}

\begin{figure*}
    \centering
     \includegraphics[scale=0.40]{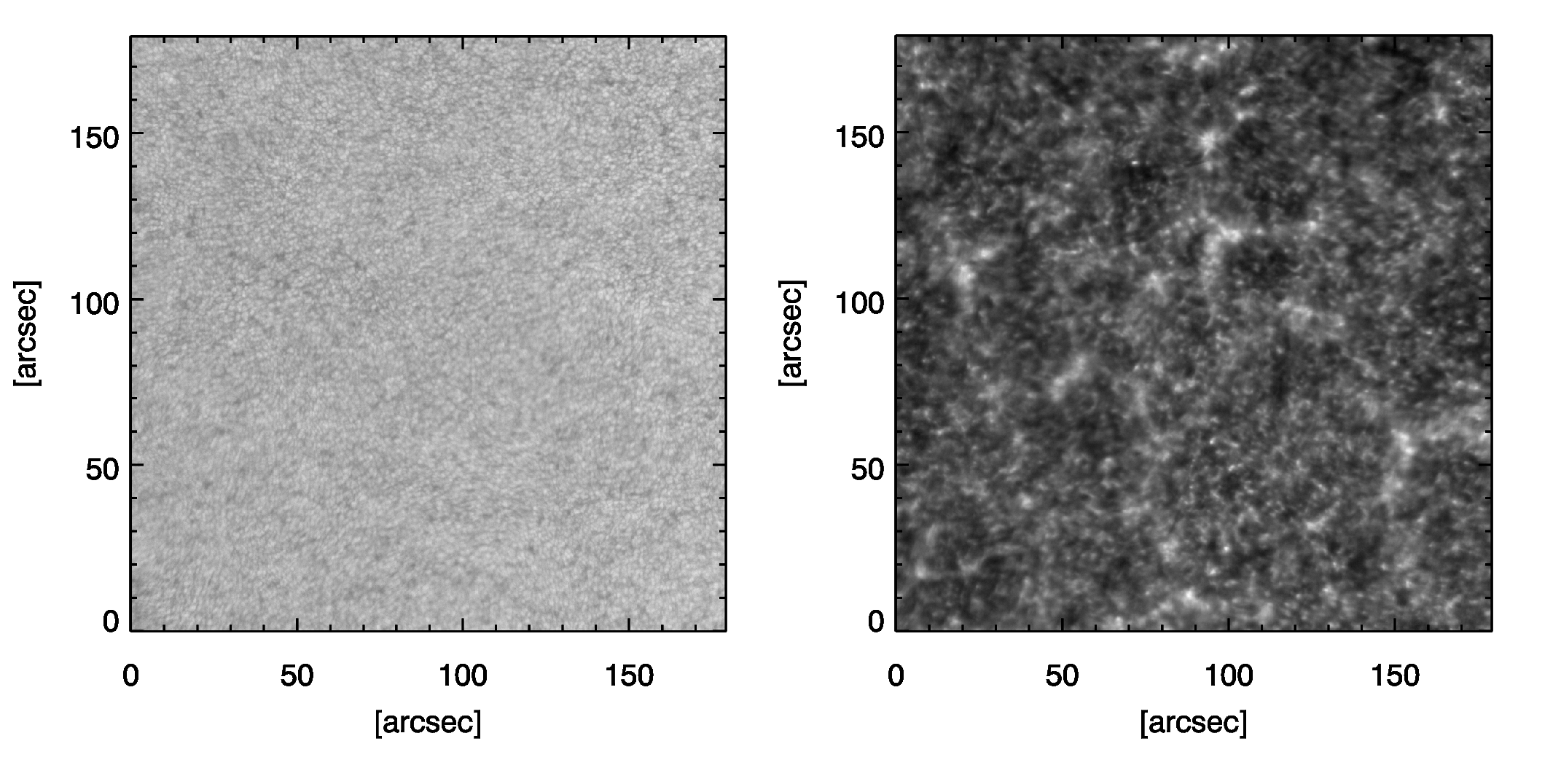}
        \caption{Sample maps of dark and flat corrected photospheric continuum intensity (\textit{left panel}) obtained from the photospheric Fe I line scan and chromospheric Ca II line core intensity (\textit{right panel}), respectively, as observed with NBI/MAST for a quiet Sun magnetic network region.}
    \label{Figure1}
\end{figure*}

The photospheric and chromospheric line-scan observations were obtained from the MAST, which is a 50-cm off-axis Gregorian Solar telescope capable of providing near-simultaneous observations of the photosphere and chromosphere of the Sun. The maximum field-of-view (FOV) of the MAST is 6 arcmin, and the theoretical diffraction limit of the telescope at 5000 {\AA} wavelength is 0.252 arcsec. To obtain near-simultaneous observations, an imager optimized with two or more wavelengths is integrated with this telescope. The Narrow Band Imager (NBI: \cite{2014SoPh..289.4007R, 2017SoPh..292..106M}) has been developed with the combination of two voltage-tunable lithium niobate Fabry-Perot etalons along with a set of interference-blocking filters. These etalons are used in tandem for photospheric observations in the Fe I 6173 {\AA} (hereafter Fe I) line. However, only one of the etalons is used for the chromospheric observations in the Ca II 8542 {\AA} (hereafter Ca II) line. Utilizing the capability of the NBI instrument with the MAST, we observed a quiet-magnetic network region located in the solar disc centre on January 02, 2023, from 05:39:54 UT to 07:32:52 UT duration. The Fe I line has been scanned at 35 wavelength positions at an equal sampling of 10 m{\AA} within 25 s, and there is a delay of 21 s between two consecutive Fe I line scan observations. However, the Ca II line has been scanned at 27 wavelength positions on both sides from the centre of the line profile at [-750.0, -610.0, -510.0, -420.0, -340.0, -270.0, -210.0, -160.0, -120.0, -85.0, -55.0, -30.0, -10.0, 0.0, 10.0, 30.0, 55.0, 85.0, 120.0, 160.0, 210.0, 270.0, 340.0, 420.0, 510.0, 610.0, 750.0] m{\AA} positions within 40 s, and there is a delay of 6 s between two consecutive Ca II line scan observations. Combining the time delay between two consecutive line scans, we have the Fe I and Ca II observations together at a cadence of 46 s. Figure \ref{Figure1} represents the sample dark and flat corrected photospheric intensity image in Fe I observations near the continuum and chromospheric line core intensity image from Ca II line scan observations. The right panel of Figure \ref{Figure1} clearly shows the significant presence of quiet regions as well as magnetic network and internetwork regions. The left panel of Figure \ref{Figure3} shows the Fe I line profile constructed from the average over full FOV and black triangles denote the locations where the line profile has been scanned, whereas the right panel of Figure \ref{Figure3} shows the Ca II line profile and the triangles depict the locations where line profile has been scanned.\\

\subsection{SDO Data}

The observed FOV also contains magnetic network regions where the magnetic field strength is of the order of kG \citep{1993SSRv...63....1S}. To get the information on magnetic fields in the network regions, we have utilized photospheric line-of-sight magnetograms obtained from the Helioseismic Magnetic Imager (HMI: \cite{2012SoPh..275..229S}) instrument onboard the SDO spacecraft, which observes the photosphere of the Sun in Fe I 6173 {\AA} line. The HMI instrument provides photospheric observables: continuum intensity, line-depth, line width, line-of-sight Dopplergrams, and line-of-sight magnetograms at a temporal cadence of 45 s with a plate scale of 0.504 arcsec per pixel.

\begin{figure}
    \centering
    \includegraphics[scale=0.25]{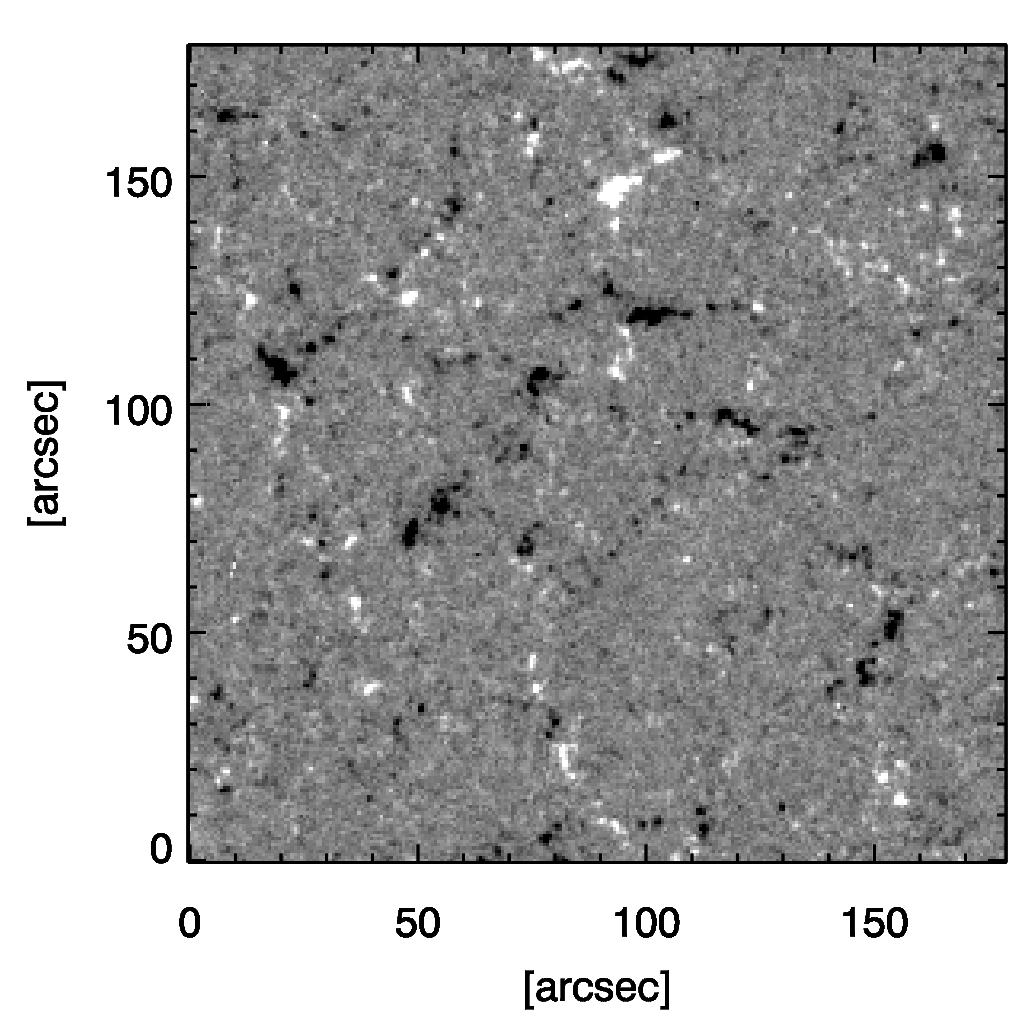}
    \caption{Sample map of photospheric line-of-sight magnetic fields obtained from the HMI/SDO spacecraft at the beginning of the observations of the same region as shown in Figure \ref{Figure1}. It has been saturated between $\pm$ 50 G to enhance small-scale magnetic features.}
    \label{Figure2}
\end{figure}

\begin{figure*}
    \centering
  \includegraphics[scale=0.30]{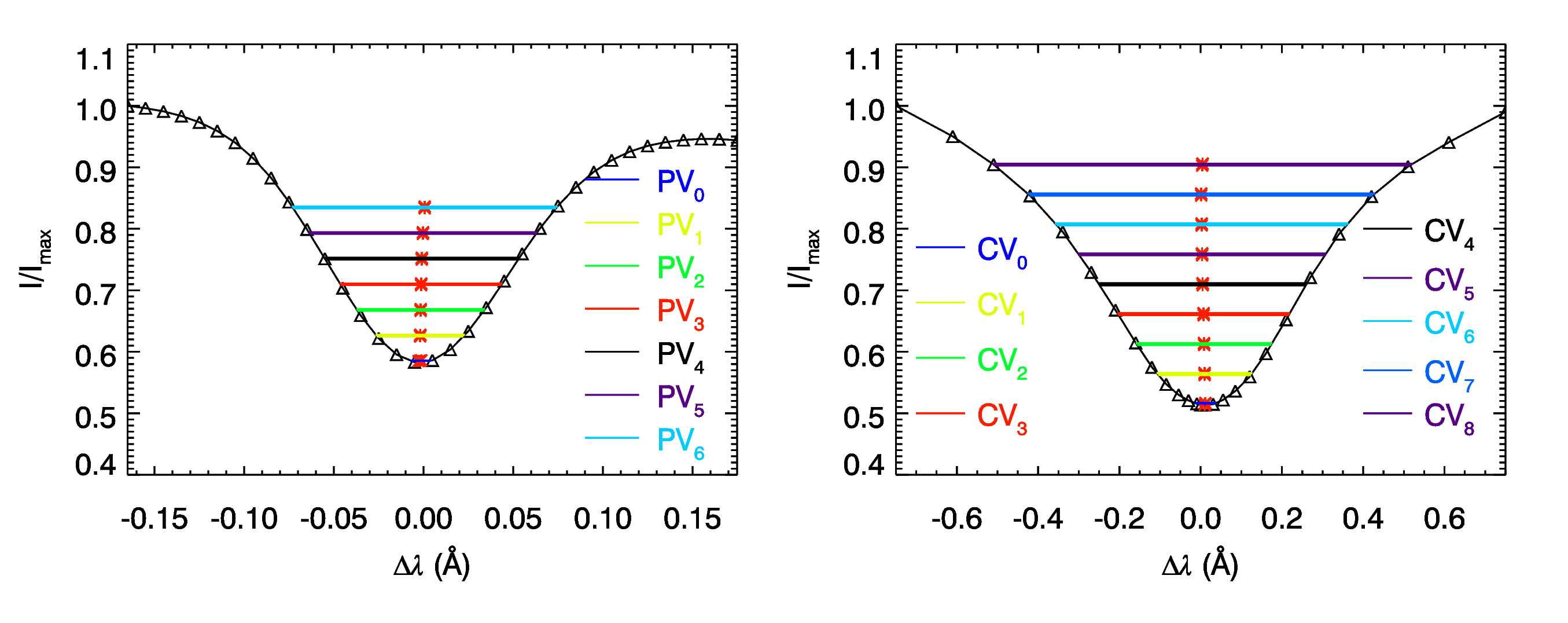}
        \caption{Sample Fe I (\textit{left panel}) and Ca II (\textit{right panel}) spectral line profiles reconstructed from the average intensity over full FOV, where triangles denote the locations where line profiles have been scanned, respectively. The horizontal lines connect the blue and red wings with bisector points.}
    \label{Figure3}
\end{figure*}

\begin{figure*}
    \centering
  \includegraphics[scale=0.40]{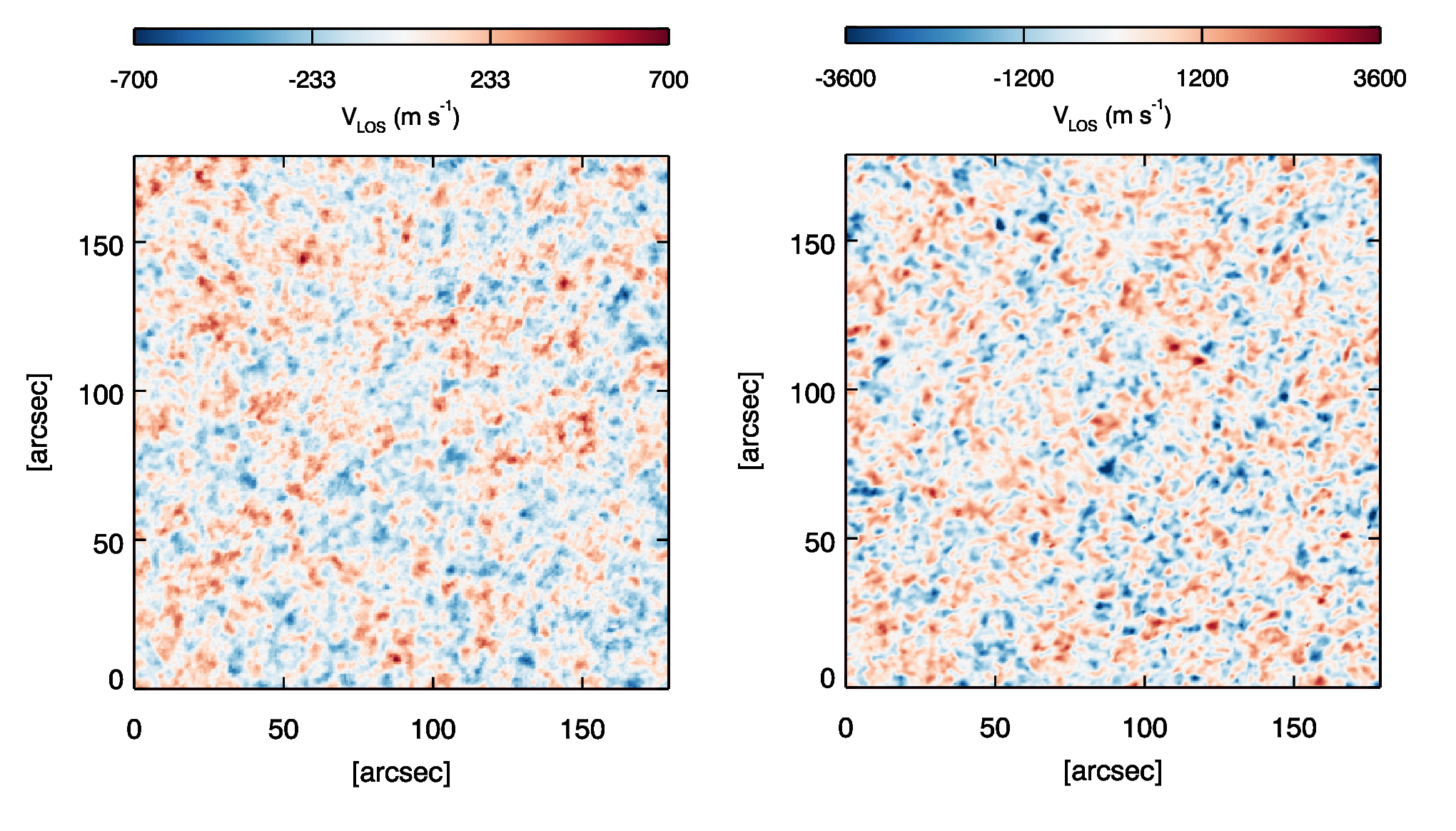}
        \caption{Sample maps of running difference of photospheric (\textit{left panel}) and chromospheric (\textit{right panel}) line-of-sight Dopplergrams at PV2 and CV2 levels in the photospheric and chromospheric line profiles, respectively, at the start time of the observations.}
    \label{Figure4}
\end{figure*}

\begin{figure}
    \centering
  \includegraphics[scale=0.25]{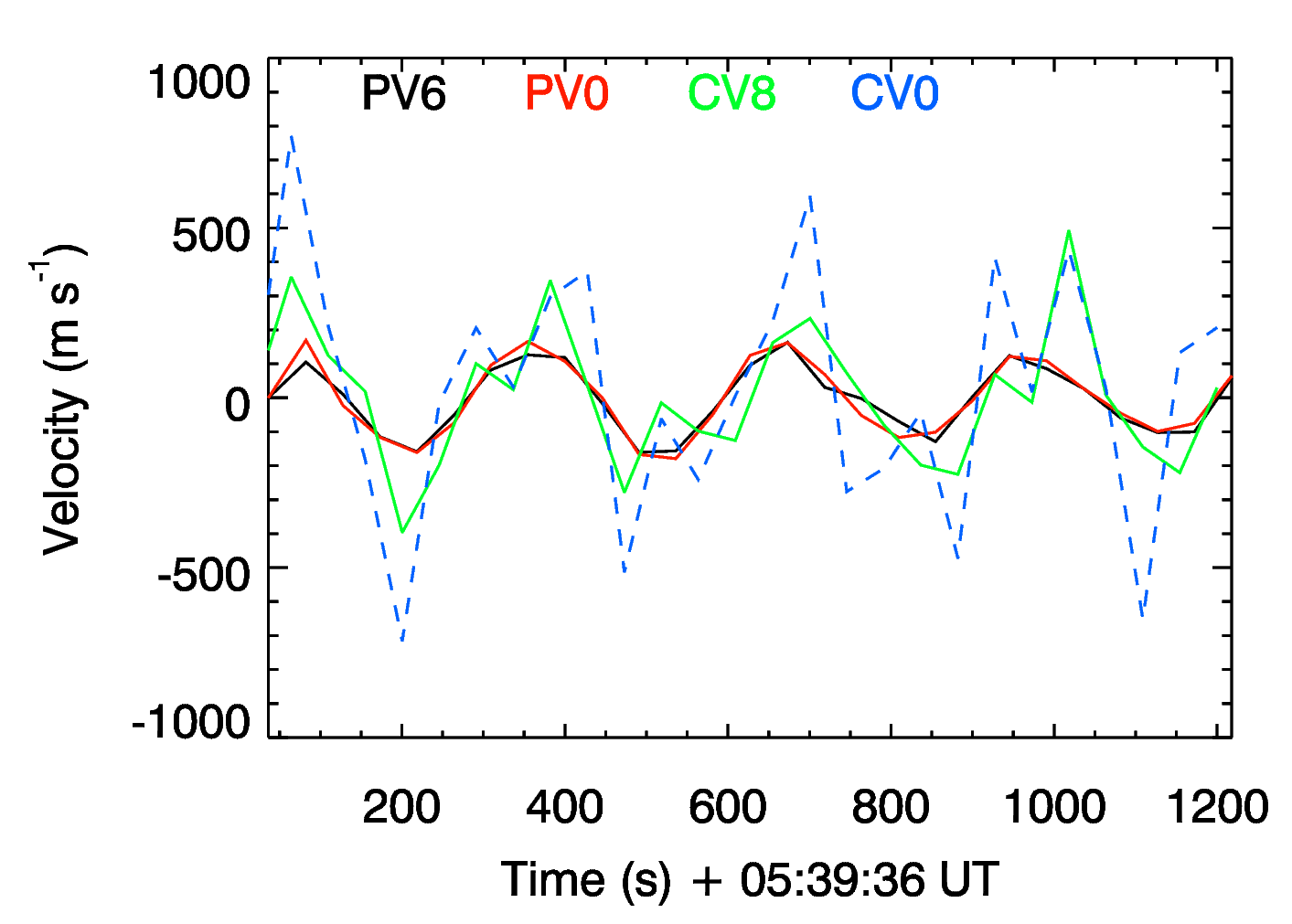}
    \caption{Velocity oscillations averaged over 20$\times$20 pixels in the middle of FOV as shown in Figure \ref{Figure1} from the photospheric to the chromospheric layers for an initial 20 min duration of the observations, showing the similarity between these oscillations at various layers in the lower solar atmosphere.}
    \label{Figure5}
\end{figure}

\section{Analysis and Results}

In order to measure the phase shift of low-frequency acoustic (evanescent) waves in the quiet Sun, we essentially require multi-height velocity observations covering the photospheric-chromospheric layers of the Sun. The bisector method \citep{Gray1976} is considered to be a fast and simple tool to infer line-of-sight velocity at multiple heights from a single line scan observations. \cite{2020A&A...634A..19G} have utilized the bisector method on photospheric synthetic Si I 10827 {\AA} line profile in the quiet-Sun and estimated multi-height velocities at different intensities. They compared it with the simulated velocity derived at various optical depths and reported good correlations with the velocity estimated at different heights in the solar atmosphere. \cite{2020ApJ...891..119B}, and \cite{2020ApJ...902...30B} used the bisector method to infer line-of-sight velocity at different line depths on the Ca II 8542 {\AA} and H$_{\alpha}$ 6563 {\AA}  line profiles in a sunspot atmosphere to investigate the inverse Evershed flow. \cite{2022ApJ...938..143G} have used the bisector method on Ca II 8542 {\AA} line profile to estimate multi-height line-of-sight velocity in the atmosphere of magnetic pores to study the propagation of coherent waves. \cite{2023JASTP.24706071K} utilized the bisector method on the Ca II 8542 {\AA} line-scan observations to estimate line-of-sight velocity over a quiet-Sun magnetic network region to study the propagation of low-frequency magnetoacoustic waves in the aforementioned magnetized regions of the Sun. Therefore, to obtain multi-height velocities from the photosphere to the chromosphere, we have also applied the well-tested bisector method on photospheric Fe I and chromospheric Ca II spectral line profiles over a quiet-Sun magnetic network region. Before utilizing the bisector method, the line scan observations have been subjected to dark and flat corrections. The corrected line-scan intensity images are further co-aligned with their next line-scan counterparts and are subjected to the pre-filter correction. In order to do the pre-filter correction, we have normalized the Fe I and Ca II line scan profiles with the available pre-filter profile. The pre-filter profile is clipped between the measured line scan wavelength range by extracting the central wavelength of the pre-filter. To extract the central wavelength of the pre-filter, observed profiles and the profile obtained from the BASS2000 (\url{https://bass2000.obspm.fr/home.php}) spectrum are compared by shifting the pre-filter profile to get the same order of change at the wings. The bisector method was used on the pre-filter corrected co-aligned line-scan intensities, and thus we obtained maps of $\delta \lambda$ by taking an average of 4 bisector points at a single intensity level (approximately to different geometrical height). Following this, we have estimated $\delta \lambda$ cubes ($\delta \lambda(x,y,n)$) at seven (n = 7) heights in the Fe I line and nine (n = 9) heights in the Ca II line at equal intensity intervals as shown in the left and right panels of Figure \ref{Figure3}, respectively. The red asterisk denotes the centre of the horizontal lines connecting the blue and red wings of the spectral line profiles. Further, we have also corrected the field-dependent wavelength shift (centre to edge). For this purpose, we have used a diffuser to take the line-scan observations, and these images are further subjected to dark and flat corrections. Further, utilizing the bisector method, we obtain a map of $\delta \lambda_{shift}$. This map was subtracted from the $\delta \lambda$ cubes, and finally we got the maps of line-of-sight velocity ($v_{LOS}$) using the expression; $v_{LOS} = \frac{\delta \lambda}{\lambda_{0}} c$, where $\lambda_{0}$ is the rest wavelength, whereas $c$ is the speed of light. \cite{2006SoPh..239...69N} has examined the formation height of the Fe I line profile using the VAL-C model \citep{1981ApJS...45..635V} and suggested that continuum forms around 16 km, whereas line core forms around 302 km above $\tau_{c}$=1 region. Moreover, the core of the Ca II line forms around 1300 km in the middle chromosphere, whereas filtergrams at -600 m{\AA} sample the photosphere at a height of approximately 200 km \citep{2014ApJ...781..126O}. Thus, we believe that after estimating seven velocity pairs in the Fe I line starting from PV$_{6}$ to PV$_{0}$, where PV$_{6}$ samples lower height and PV$_{0}$ samples upper height in the Fe I spectral line profile, respectively, and nine bisector velocity pairs in Ca II line ranging from CV$_{8}$ to CV$_{0}$; where CV$_{8}$ samples lower height and CV$_{0}$ samples upper height in the Ca II spectral line profile, respectively, approximately we have multi-height line-of-sight velocities covering from the photosphere to the middle chromosphere of the Sun. Figure \ref{Figure4} shows the sample maps of line-of-sight velocities at PV2 and CV2 pair after taking the running difference of Dopplergrams. We have taken running differences of Dopplergrams to remove slowly varying signals. Sample velocity oscillations are depicted in Figure \ref{Figure5} ranging from the lowest height in the Fe I line profile to the core of the Ca II line profile, which has been averaged over a FOV of 20$\times$20 pixels$^2$, indicating similarities between oscillations at various heights in the lower solar atmosphere. \\

\textbf{\subsection{Phase shift and phase travel time within the  photospheric layers}}

In order to estimate the phase difference between two height velocity--velocity pairs from these multi-height velocities, first of all, we have interpolated all the velocity signals to a common temporal cadence of 45 s incorporating the time lag present in the start time of the observations and used the following expression to estimate the cross-spectra.\\

The phase shift and coherence between two evenly sampled time series ($A$ and $B$) at each pixel over the full FOV have been estimated in the following way:

\begin{equation}
X_{AB}(\nu) =  I_{A}(\nu)\times I^{*}_{B}(\nu) 
\end{equation}

Where I's are the Fourier transforms, and $*$ denotes the complex conjugate. The phase difference between two-time series is estimated from the phase of complex cross product ($X_{AB}(\nu)$).\\

\begin{equation}
\delta\phi(\nu) = tan^{-1}(Im(<X_{AB}(\nu)>)R(<X_{AB}(\nu)>)).
\label{deltaphase}
\end{equation}  

Here, positive $\delta\phi(\nu)$ means that signal 1 lead\textbf{s} 2, i.e., a wave is propagating from lower height to upper height and vice-versa, and the magnitude of $X_{AB}(\nu)$ is used to estimate the coherence (C), which ranges between 0 to 1. Coherence is a measurement of the linear correlation between two signals, where 0 indicates no correlation, and 1 means perfect correlation. It can be estimated using the following expression: 

\begin{equation}
C(\nu) = sqrt(|<X_{AB}(\nu)>|^{2} / <|I_{A}|^{2}> <|I_{B}|^{2}>).
\label{cohere}
\end{equation}

\begin{figure}
    \centering
  \includegraphics[scale=0.30]{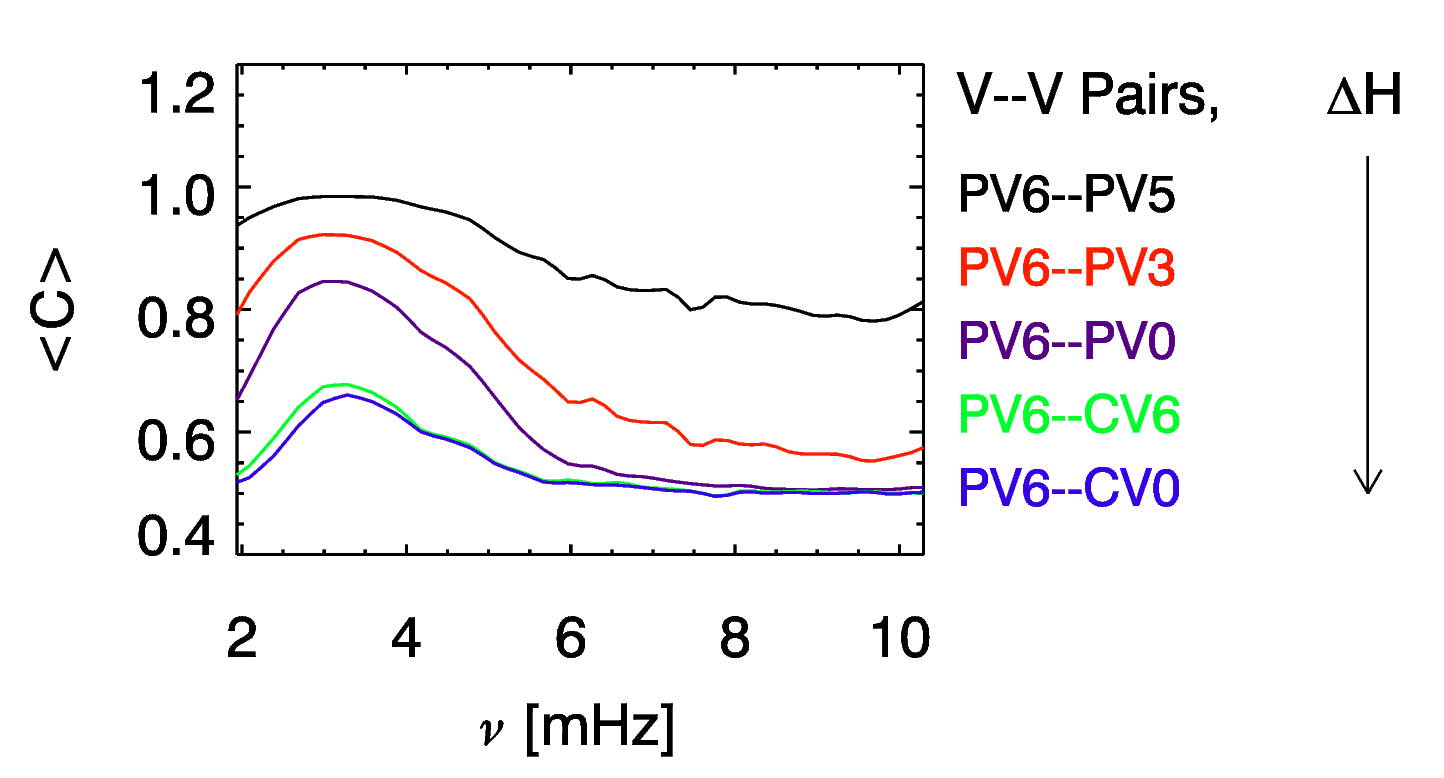}
    \caption{Average coherence over the full FOV estimated from two height velocity--velocity pairs from lowest height difference to highest height difference range, clearly depicting the decreasing coherence as height difference ($\Delta H$) increases.}
    \label{Figure6}
\end{figure}

\begin{figure*}
    \centering
  \includegraphics[scale=0.30]{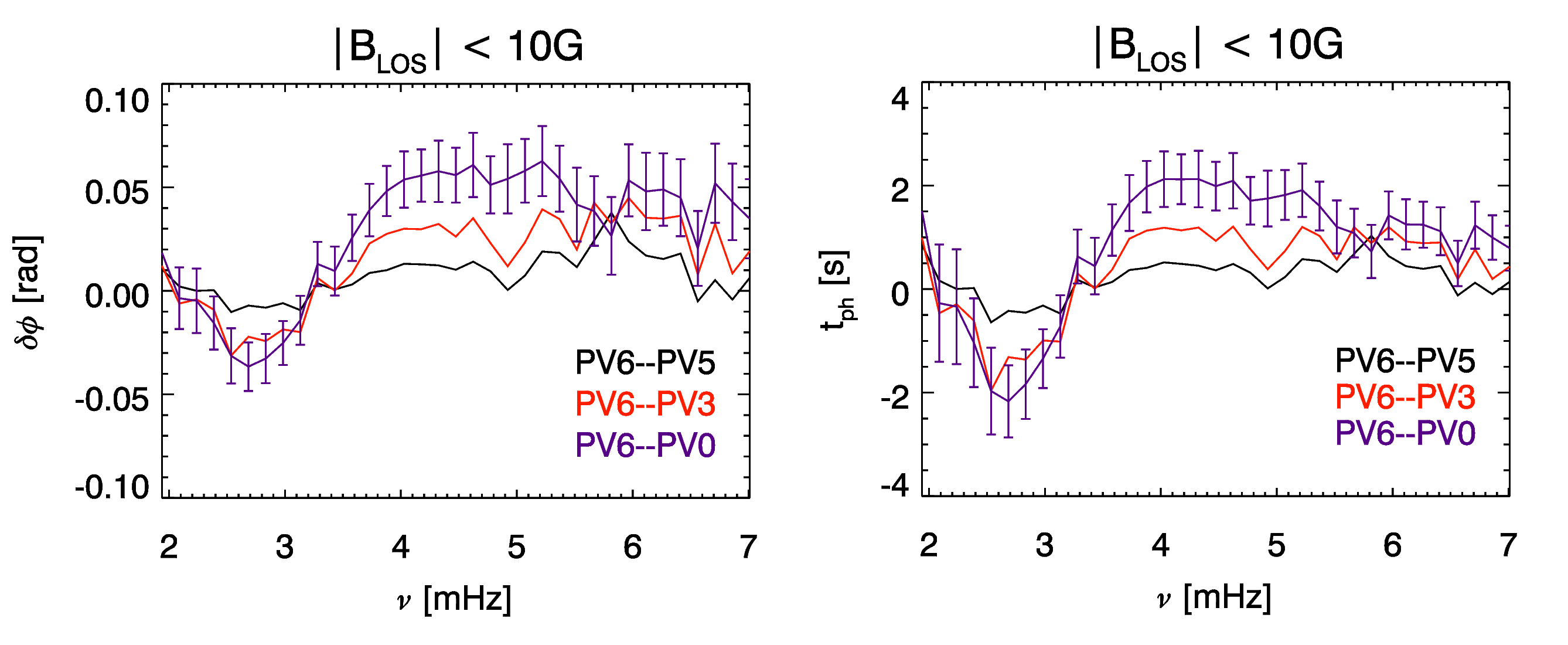}
        \caption{Phase shifts measured from PV5, PV3 and PV0 relative to PV6, displayed as a function of frequency (\textit{left panel}) integrated over the pixels having $|B_{LOS}|<$ 10 G and coherence greater than 0.5 over the full FOV. Standard deviations are displayed only for one of these curves, while the errors for other curves are similar. Right panel: Same as \textit{left panel}, but relative time shifts are displayed.}
    \label{Figure7}
\end{figure*}

\begin{figure*}
    \centering
  \includegraphics[scale=0.30]{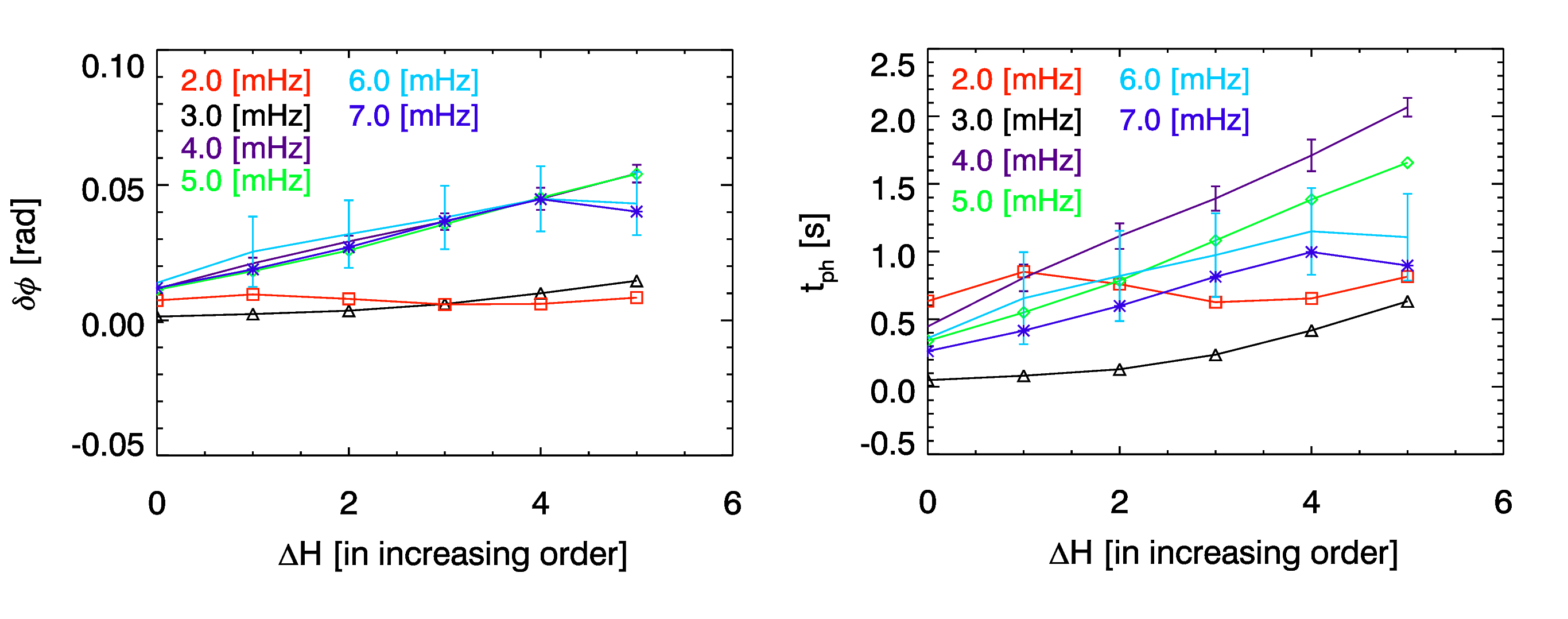}
        \caption{Relative phase shifts within the photospheric height range (\textit{left panel}) from PV6-PV5, PV6-PV4, PV6-PV3, PV6-PV2, PV6-PV1, PV6-PV0, two height V-V pairs, displayed as a function of height difference ($\Delta H$) for different frequency bands. Standard deviations are displayed only for the 4.0 and 6.0 mHz frequency bands, while the errors for other curves are within the plotted error bars. Right panel: Same as left panel, but the relative time shifts are displayed.}
    \label{Figure8}
\end{figure*}

Before estimating the phase difference $\delta\phi(\nu)$ between two height V--V pairs, we have estimated the coherence using Equation \ref{cohere} between various two height V--V pairs. Figure \ref{Figure6} represents mean coherence over the full FOV estimated between two height velocities as indicated within the plots. We observe that as the height difference ($\Delta H$) between two height velocity pairs is small, estimated coherence is very high (c.f., PV6--PV5, V--V pair), while coherence decreases as the $\Delta H$ between two height V--V pairs increases (c.f., PV6--CV0, V--V pair). The main focus of the analysis is on the low-frequency acoustic waves i.e. $2-5$ mHz band, in which we found coherence is greater than 0.5. The phase shift estimated between various two height V--V pairs within the photospheric height range is shown in the left panel of Figure \ref{Figure7}. We have selected only those pixels where the absolute line-of-sight magnetic field strength is less than 10 G to rule out the effect of magnetic fields on the phase shift measurements. Magnetized pixels excluded with this criterion occupy about 26$\%$ area within the full FOV being used in our analysis. Further, only those pixels have been considered which depict coherence greater than 0.5. Here, we observe that $\delta \phi$ is not only a function of $\nu$ but also a function of $\Delta H$ between two height V--V pairs (c.f., left panel of Figure \ref{Figure7}), clearly indicating that as $\Delta H$ increases phase shift is also increasing. For the upward propagating waves, it is usually expected that phase shift is small for neighbouring layers while higher for increasing height difference layers. However, for the non-evanescent waves, it should be zero or negligible. We also calculate phase travel time from the phase difference as follows, $t_{ph} = \frac{\delta \phi}{2 \pi \nu}$, which gives information about the time lag corresponding to phase shift. The right panel of Figure \ref{Figure7} shows the travel time estimated within the quiet-Sun showing $t_{ph}$ around 1 s at 2 mHz, nearly -2 s at 2.5 mHz 0 s around 3.2 mHz and then increases up to 2 s above 3 mHz and then again starts decreasing around 5 mHz. It is noteworthy that a change in sign around 2 mHz and 3.2 mHz is clearly observed. 

Furthermore, to better understand the height evolution of phase shift and phase travel time, we have also examined phase shift and phase travel time as a function of $\Delta H$ estimated at 2, 3, 4, 5, 6, and 7 mHz, which is integrated over a certain frequency bin i.e. $\pm$ 0.25 mHz from the centre of the mentioned frequencies. Figure \ref{Figure8} shows the phase difference and phase travel time as a function of the $\Delta H$ between the photospheric height range. It is generally expected that phase shift and phase travel time should be zero or negligible for the frequencies ($\nu$) $<$ 5.0 mHz for the evanescent waves. However, we note that the phase difference and phase travel time at 3 mHz is close to 0, while at 2 mHz, the phase difference is rather having a very small value, which gives non-zero phase travel time. On the other hand, phase difference and phase travel time at 4, 5, 6, and 7 mHz increases as a function of $\Delta H$ within the photosphere.

\textbf{\subsection{Phase shift and phase travel time within the photosphere-chromosphere interface regions}}

\begin{figure*}
    \centering
  \includegraphics[scale=0.30]{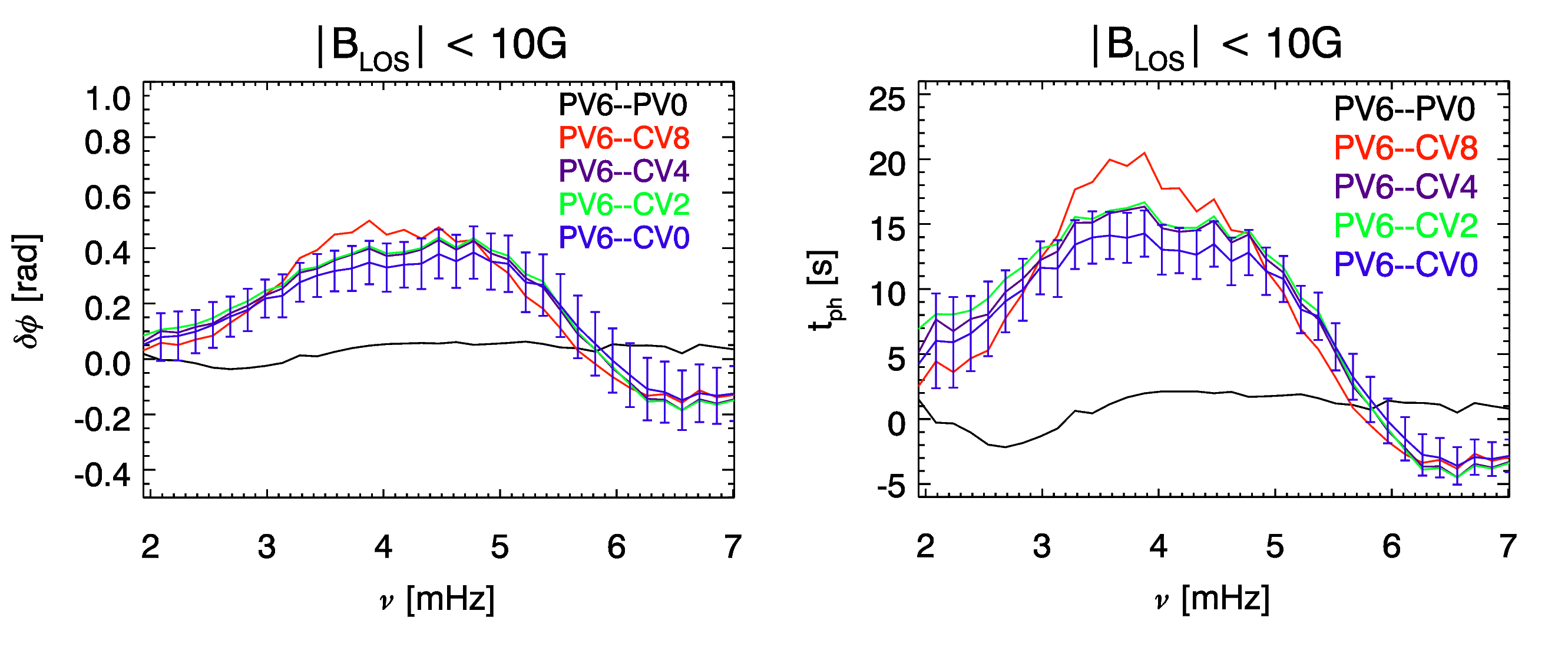}
        \caption{Phase shifts measured from PV0, CV8, CV4, CV2, and CV0 relative to PV6, displayed as a function of frequency (\textit{left panel}) integrated over the pixels having $|B_{LOS}|<$ 10 G and coherence greater than 0.5 over the full FOV. Standard deviations are displayed only for one of these curves, while the errors for other curves are similar. Right panel: Same as \textit{left panel}, but the relative time shifts are displayed.}
    \label{Figure9}
\end{figure*}

\begin{figure*}
    \centering
  \includegraphics[scale=0.30]{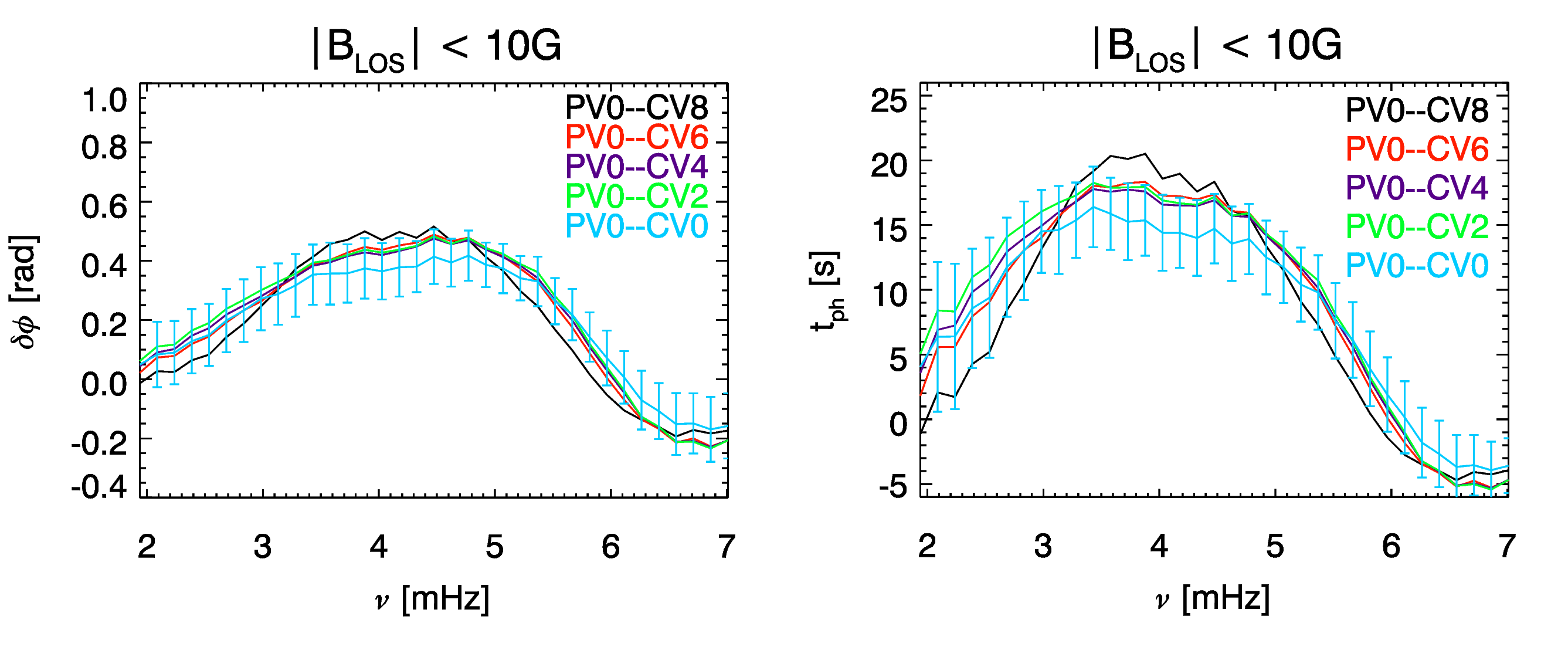}
        \caption{Phase shifts measured from CV8, CV6, CV4, CV2 and CV0 relative to PV0, displayed as a function of frequency (\textit{left panel}) integrated over the pixels having $|B_{LOS}|<$ 10 G and coherence greater than 0.5 over the full FOV. Standard deviations are displayed only for one of these curves, while the errors for other curves are similar. Right panel: Same as \textit{left panel}, but the relative time shifts are displayed.}
    \label{Figure10}
\end{figure*}

Utilizing the multi-height velocities estimated within the Ca II line along with velocities derived from the Fe I line scan, we estimate phase difference and phase travel time within the photosphere-chromosphere interface regions, which are shown in Figure \ref{Figure9}. Interestingly, we note that phase difference and phase travel time estimated from the combination of photospheric and chromospheric velocity pairs, as illustrated in the plots, indicate that chromospheric oscillations significantly lag photospheric oscillations in $2-5$ mHz band, and at around 5 mHz phase difference and phase travel time start decreasing, showing a change in sign around 6 mHz, whereas within the photosphere it occurs around 3.2 mHz. We also notice that when the $\Delta H$ between photospheric and chromospheric two height V-V pairs is small i.e. PV6-CV8, as shown in Figure \ref{Figure9} by red colour, the plot depicts that broadening is small and peaks around 4 mHz as compared to the higher height difference photospheric and chromospheric velocity pairs i.e. PV6-CV0, which shows a slight broadening compared to PV6-CV8 velocity pairs, because higher height difference velocities pairs receive higher frequency contributions. Nevertheless, we also observe that a small amount of change in phase difference gives a significant amount of phase travel time as shown by the red colour plot in the right panel of Figure \ref{Figure9}. We have also estimated phase difference and phase travel time by using photospheric PV0 velocity as the lowest height velocity and all chromospheric velocities, shown in Figure \ref{Figure10}. We observe that the phase difference and phase travel time are almost similar to the phase difference and phase travel time estimated from the PV6 and all chromospheric velocity pairs as depicted in Figure \ref{Figure9} with some visual differences. \\

\section{Discussion and Conclusions}

In this article, we present an analysis of the phase shift and phase travel time of acoustic waves in the quiet-Sun located in the solar disc centre, focusing within the photosphere and photosphere-chromosphere interface regions of the Sun. Our investigation utilizes multi-height velocities estimated by employing the bisector method within the Fe I and Ca II line scan observations obtained from the NBI/MAST. Using multi-height velocities, our analysis shows a non-zero phase shift of low-frequency acoustic waves in the quiet Sun. Notably, a change in sign around 3 mHz is observed in the phase difference measured within the photospheric velocity pairs, indicating both evanescent and propagating characteristics of low-frequency ($\nu$ $<$ 5.0 mHz) acoustic waves. 

Recently, \cite{2022ApJ...933..109Z} utilizing high-resolution observations of the photosphere in the Fe I spectral line obtained from the IBIS/DST, found a similar non-zero phase shift and phase travel time of evanescent waves in the quiet-Sun region surrounding a small sunspot. They also found a change in sign around 3 mHz. \cite{2023ApJ...949...99W} utilizing 3D simulations and from the synthesis of Fe I line also found non-zero phase shift of evanescent waves. However, the change in sign in the simulation was observed around 4 mHz. Contrary to the suggestions of \cite{2018ApJ...853..161C} that the center-to-limb effect is also responsible for the change in sign, which occurs around 5.5 mHz, our study and \cite{2022ApJ...933..109Z} observed the change in sign around 3 mHz. Nevertheless, in our analysis exclusion of pixels with absolute line-of-sight magnetic fields greater than 10 G in the quiet-Sun observations yielded results consistent with \cite{2022ApJ...933..109Z}, reinforcing the notion that the observed change in sign is not influenced by moat flows or sparse magnetic fields present surrounding the sunspot as pointed out by \cite{2022ApJ...933..109Z}. Despite differences in spatial resolutions and modulation transfer functions between our observations using NBI/MAST and \cite{2022ApJ...933..109Z} using IBIS/DST, we obtained similar results. This finding suggests that the change in sign at 3 mHz is likely due to some physical phenomenon rather than observational or instrumental factors. The most promising explanation of this non-zero evanescent nature of low-frequency acoustic waves comes from the fact that the Fe I line has asymmetry. Therefore, the velocity derived at various intensity levels within the line itself may lead to a non-zero phase shift. Another strong possibility is the non-adiabatic nature of the solar atmosphere, as suggested by \cite{2022ApJ...933..109Z} and \cite{2023ApJ...949...99W}. For evanescent waves, where phase changes in velocities may be subtle, the temperature responses and observed intensity changes at different heights may not occur simultaneously due to height-dependent time delays resulting from the nonadiabatic nature of the atmosphere. Overall, the complicated interplay between the nonadiabatic properties of the atmosphere and the spectral line's characteristics influences height- and frequency-dependent phase shifts in Doppler velocities.\\

Additionally, our analysis incorporated chromospheric Ca II line scans, revealing a decrease in phase shift around 5 mHz and a change in sign in the 5.5-5.8 mHz range. We speculate that the decrease in phase shift is possibly due to the refraction and downward propagation of high-frequency acoustic waves from the higher solar atmospheric layer. The inclusion of different photospheric height velocities and chromospheric velocities further demonstrates the height and frequency-dependent variation in the occurrence of the change in sign of the phase difference of the acoustic waves in the solar atmosphere. This finding aligns with the trends observed by \cite{2022ApJ...933..109Z}, where they have used line core and continuum intensity observations obtained from the photospheric Fe I 6173 {\AA}  line from the IBIS/DST instrument.\\

In conclusion, our comprehensive analysis supports the existence of a non-zero phase shift and phase travel time of low-frequency acoustic waves in the quiet-Sun region, with the observed change in sign at 3 mHz within the photosphere, attributed to physical phenomena rather than observational or instrumental effects. The inclusion of chromospheric data and consideration of height differences in velocity pairs contribute to a refined understanding of the frequency-dependent characteristics of these waves. In future, multiline inversion of simultaneous spectropolarimetric observations of the photosphere and the chromosphere would be beneficial to further understand height-dependent phase shift variations in the quiet Sun.


\begin{acknowledgments}
This work utilized data obtained from the 50 cm aperture Multi Application Solar Telescope operated by the Udaipur Solar Observatory, Physical Research Laboratory, Dept. of Space, Govt. of India. We are thankful to the MAST team for providing the data used in this work. We acknowledge the use of data from the HMI instrument onboard the Solar Dynamics Observatory spacecraft of NASA. We are thankful to the SDO team for their open data policy. The work carried out at Udaipur Solar Observatory, Physical Research Laboratory, Udaipur, India, is supported by the Dept. of Space, Govt. of India. We are thankful to the reviewer for the suggestions that improved the presentation of results as well as the inclusion of relevant discussions in our manuscript.
\end{acknowledgments}

\bibliography{HK_mans}{}
\bibliographystyle{aasjournal}



\end{document}